\documentclass[superscriptaddress,aps,pra,amssymb,amsfonts]{revtex4}

\usepackage{amsmath,amsthm}
\input{Qcircuit}

\usepackage{graphicx,color,graphics}

\newtheorem{theorem}{Theorem}

\def\one{{\mathchoice {\rm 1\mskip-4mu l} {\rm 1\mskip-4mu l} {\rm
1\mskip-4.5mu l} {\rm 1\mskip-5mu l}}}

\newcommand{\braket}[2]{\langle #1 | #2 \rangle}

\newcommand{\sket}[1]{|#1\rangle }

\def\mP{{\mathfrak P}}

\def\bV{{\bar V}}

\def\pp{{\parallel}}

\begin{document}

\title{Efficient discrete-time simulations of continuous-time quantum query algorithms}

\author{R. Cleve}
\email{cleve@cs.uwaterloo.ca}
\affiliation{Institute for Quantum Computing, University of Waterloo, Waterloo, ON N2L 3G1, Canada}
\affiliation{Perimeter Institute for Theoretical Physics, Waterloo, ON N2L 2Y5, Canada}

\author{D. Gottesman}
\email{dgottesman@perimeterinstitute.ca}
\affiliation{Perimeter Institute for Theoretical Physics, Waterloo, ON N2L 2Y5, Canada}

 \author{M. Mosca}
 \email{mmosca@iqc.ca}
\affiliation{Institute for Quantum Computing, University of Waterloo, Waterloo, ON N2L 3G1, Canada}
\affiliation{Perimeter Institute for Theoretical Physics, Waterloo, ON N2L 2Y5, Canada}

\author{R. D. Somma}
\email{rsomma@perimeterinstitute.ca}
\affiliation{Perimeter Institute for Theoretical Physics, Waterloo, ON N2L 2Y5, Canada}

\author{D. L. Yonge-Mallo}
\email{davinci@iqc.ca}
\affiliation{Institute for Quantum Computing, University of Waterloo, Waterloo, ON N2L 3G1, Canada}

\date{\today}

\begin{abstract}
The continuous-time query model is a
variant of the discrete query model in which queries can be interleaved with
known operations (called ``driving operations") continuously in time.
Interesting algorithms have been discovered in this model, such as an
algorithm for evaluating {\sc nand} trees more efficiently than any classical
algorithm. Subsequent work has shown that there also exists an efficient
algorithm for {\sc nand} trees in the discrete query model; however, there is
no efficient conversion known for continuous-time query algorithms for
arbitrary problems.

We show that any quantum algorithm in
the continuous-time query model whose total query time is $T$ can be simulated
by a quantum algorithm in the discrete query model that makes
$O(T \log T /\log\log T) \subset \tilde{O}(T)$
queries. This is the first upper bound that is independent of the driving
operations (i.e., it holds even if the norm of the driving Hamiltonian is very
large). A corollary is that any lower bound of $T$ queries for a problem in the
discrete-time query model immediately carries over to a lower bound of
$\Omega(T \log \log T/\log T) \subset \tilde{\Omega}(T)$ in the continuous-time query model.
\end{abstract}

\maketitle

\section{Introduction and Summary of Results}

In the \textit{query} (a.k.a.~black-box or oracle) model of computation, one is given a
black box that computes the individual entries of an $N$-tuple,
$x =(x_0, x_1, \ldots, x_{N-1})$,
and the goal is to compute some function of these values, making as few queries
to the black-box as possible.
Many quantum algorithms can be naturally viewed as algorithms in this model, including
Shor's factoring algorithm~\cite{Shor1997}, whose primary component computes the
periodicity of a periodic sequence $x_0, x_1, \ldots, x_{N-1}$ (technically,
the sequence must be also be distinct within each period).
Other examples are~\cite{Grover1997,Ambainis2007,BrassardH+1997}.

In the quantum query model, a (full) quantum query is a unitary operation $Q_x$
such that
\begin{equation}\label{eq:discrete1}
Q_x \ket{j}\ket{b} = \ket{j}\ket{b \oplus x_j},
\end{equation}
for all $j \in \{0,1,\ldots,N\!-\!1\}$ and $b$ from the set of values that entries of
the $N$-tuple ranges over, and $\oplus$ can be set to the bit-wise exclusive-or.
Queries are interleaved with other quantum operations that ``drive'' the computation.
The \textit{query cost} of an algorithm is the number of queries that it makes.
The efficiency of the other operations, besides queries, is also of interest.
An algorithm is deemed efficient if it is efficient in both counts.

When the tuple $x$ consists of binary values, the form of a full query
can be equivalently expressed as
\begin{equation}\label{eq:discrete2}
Q_x \ket{j}\ket{b} = (-1)^{b \cdot x_j}\ket{j}\ket{b},
\end{equation}
which is related to $Q_x$ from Eq.~(\ref{eq:discrete1}) via conjugation with a Hadamard
transformation on the second register.
For convenience of notation, we can absorb the second qubit register $b$
into the definition of $x$, by defining $x' = (x'_0, \ldots, x'_{2N-1})$
as $x'_{j0} = 0$ and $x'_{j1}= x_j$.
Henceforth, we simply omit the parameter~$b$, and define a discrete query $Q_x$
as
\begin{equation}\label{eq:discrete3}
Q_x \ket{j} = (-1)^{x_j}\ket{j},
\end{equation}
for all $j \in \{0,1,\ldots,N\!-\!1\}$.
(See~\cite{KayeL+2007} for more information about relationships between
different forms of queries.)

Farhi and Gutmann~\cite{FarhiG1998a} introduced a continuous-time variant of the
query model, where queries are performed continuously in time in the following
sense.
A \textit{query Hamiltonian} $H_x$ is defined as
\begin{equation}\label{abs:eq:cont-query}
H_x \ket{j} = x_j\ket{j},
\end{equation}
for $j \in \{0,1,\ldots,N\!-\!1\}$.
Note that evolving under $H_x$ for time $\pi$ results in the  full discrete query
$Q_x$ of Eq.~(\ref{eq:discrete3}).
A quantum algorithm in the continuous-time query model is specified by:
a \textit{driving Hamiltonian}, $D(t)$, which is an arbitrary time-dependent
Hamiltonian; an initial state $\ket{\psi_0}$; an execution time $T>0$, and a
measurement $M$ of the final state. ($D(t)$, $\ket{\psi_0}$, $T$, and $M$, are
all functions of the input size $N$.)
The input to the algorithm is embodied by a query Hamiltonian $H_x$.
In the execution of the algorithm, the initial state $\ket{\psi_0}$ evolves
under the Hamiltonian $H_x + D(t)$ from time $t=0$ to time $t=T$.
Measurement $M$ of the resulting final state determines the output
of the algorithm.

The continuous-time query model has proven to be a useful
conceptual framework for discovering new quantum
algorithms~\cite{ChildsC+2003,FarhiG+2007}.
Many algorithms in this setting can be converted to algorithms in the more
conventional quantum query model~\cite{ChildsC+2007,AmbainisC+2007}.
However, it has not been previously shown that this can be done in general
without incurring a significant loss in query complexity.

The Suzuki-Trotter formula~\cite{Suzuki1993} can be used to approximate
a continuous-time algorithm by a sequence of full queries interleaved with
unitary operations induced by $D(t)$.
This results in simulations of cost
$O(\exp(1/\eta)\,(\| D \|\,T)^{1+\eta})$
for arbitrarily small $\eta > 0$~\cite{BerryA+2007,Childs2004}
(the result was shown for the case of time-independent $D(t)$).
Although this is ``close to linear" in cases of interest where $\| D \|$ is
bounded by a constant, if $\| D \|$ grows significantly as a function of
the input size $N$, this approach fails to yield an efficient bound.
Recent work by Childs~\cite{Childs2008} gives a simulation of cost $O(\| D \|\,T)$
that applies for all $D(t)$ that are time-independent and with the additional
property that their matrix entries are nonnegative.
This raises the question of whether a ``highly energetic" driving Hamiltonian
can result in computational speedup over the discrete query model in some scenarios.
The exponential cost in $1/\eta$ for general driving Hamiltonians $D(t)$
is also undesirable.

For some problems---such as searching for a marked item or computing the parity of
the input bits---it is already known that the continuous-time model provides no
asymptotic reduction in query cost~\cite{FarhiG1998a} (regardless of $D(t)$).
Mochon~\cite{Mochon2006} raised the question of whether this
equivalency remains valid in general: most known lower bounds only apply
to the number of full queries needed to solve a problem, leaving open the
possibility that these lower bounds could be circumvented using continuous-time
queries.  We show that this cannot happen and essentially answer Mochon's question
by showing that {\em any algorithm in the continuous-time query model whose total
query time is $T$ can be simulated by an algorithm in the quantum query model that
makes $\tilde{O}(T)$ queries}.
More specifically, we prove the following theorem:

\begin{theorem}
\label{mainthm}
Suppose we are given a continuous-time query algorithm with any driving Hamiltonian
$D(t)$ whose sup norm $\| D(t) \|$ is bounded above by any $L_1$ function with respect
to $t$. (The size
of $\|D(t)\|$ as a function of the input size $N$ does not matter.)
Then there exists a discrete-time query algorithm that makes
\begin{equation}
O\left(\frac{T \log (T/\varepsilon)}{\varepsilon\log\log(T/\varepsilon)}\right)
\end{equation}
full queries and whose answer has fidelity  $1-\varepsilon$ with the output
of the continuous-time algorithm.  If the continuous-time query algorithm acts
on a classical input (for instance, if it computes a function $f(x)$ of the oracle $x$),
the discrete-time query algorithm makes
\begin{equation}
O\left(\frac{T \log (T/\varepsilon) \log (1/\varepsilon)}{\log\log(T/\varepsilon)}\right)
\end{equation}
full queries.
\end{theorem}

Note that this implies that any lower bound of $T$ proven for the discrete query model
automatically yields a lower bound of $\Omega(T \log\log T / \log T) \subset \tilde{\Omega}(T) $
for the continuous-time case.
In addition, any algorithm in the discrete query model using $T$ full queries can be
easily simulated by a continuous-time algorithm running for time $O(T)$.  This can be done with a
driving Hamiltonian that rapidly swaps qubits to effectively turn on and off the query
Hamiltonian.  Thus, the two models (discrete and continuous queries) are equivalent up to a
sub-logarithmic factor.

\subsection{Rough overview of the proof of Theorem~\ref{mainthm}}

Here we provide a rough sketch of the proof of Theorem~\ref{mainthm}; a
more detailed exposition is in Section~\ref{sec2}.
Starting with a continuous-time query algorithm, we apply the following sequence of transformations
to it.
\begin{description}
\item[A.] \textbf{Convert to a fractional query algorithm.}
Using a suitable Trotter-Suzuki type of approximation, the algorithm can be simulated by
interleaved executions of $D(t)$ and $H_x$ for small amounts of time.
The approximation uses about $p = \|D\|T^2/\varepsilon$ time slices, each of length $T/p$
for precision fidelity $1-\varepsilon$.
This does not readily convert into an efficient discrete-time query algorithm because
the straightforward way of simulating each $H_x$ evolution uses full discrete-time
queries, even though the time evolution is very small.
The total discrete query cost would be $O(\|D\|T^2/\varepsilon)$ (and even the reduced
exponent of $T$ resulting from a high-order Suzuki formula would not affect the dependence
on $\|D\|$, which could potentially be very large).

\item[B.] \textbf{Simulate fractional queries with low amplitude controlled discrete queries.}
We use a construction that permits each $H_x$ small-time evolution to be simulated by
a single \textit{controlled}-discrete query with control qubit in state
$\approx \sqrt{1-T/2p}\ket{0} + i \sqrt{T/2p}\ket{1}$.
This construction succeeds conditional on a subsequent measurement outcome.
The success probability is approximately $1-T/p$.  It is therefore very likely that
roughly $T$ of the $p$ fractional queries will fail, and a procedure for
correcting these post-selection failures is explained in step D below.

\item[C.] \textbf{Approximate segments of control qubits by low-Hamming-weight states.}
Part D will require us to divide the computation into segments, each involving $m$
small-time evolutions of $H_x$.  For each segment, the
collective state of the $m$ control qubits is
$\ket{\phi} \approx (\sqrt{1-T/2p}\ket{0} + i \sqrt{T/2p}\ket{1})^{\otimes m}$.
We can construct another $m$-qubit state $\ket{\phi'}$ such that
$|\braket{\phi}{\phi'}|^2 > 1 - \varepsilon/T$ and such that $\ket{\phi'}$ is
a superposition of basis states with Hamming weight only
$O(m(T/p)\log(T/\varepsilon)/\log\log(T/\varepsilon))$.
The Hamming weight of the control qubits is effectively the number of full queries
performed. By rearranging the circuit, we make this association
explicit, allowing us to truncate the circuit to deal with only the typical
case, and thus reduce the total number of full queries needed for the
segment to only $O(m(T/p)\log(T/\varepsilon)/\log\log(T/\varepsilon))$.

\item[D. Correct the post-selection errors for each segment.]
Returning to the post-selection errors in the simulation of the fractional queries,
they are corrected by dividing the computation into segments of sufficiently small
size so that:
(a) there are $O(T)$ segments to simulate;
(b) the expected number of errors per segment is $\le 1/8$.
The post-selection results for each segment reveal exactly where any errors occurred,
making it possible to ``undo" the segments and then attempt to compute them again.
This process is applied recursively, since new errors can arise during these
steps.
We show that this process only increases the expected number of segments simulated
(including those that arise from corrections) by a constant factor.
The result is $O(T)$ simulations of segments with an expected number of
discrete queries $O(T\log(T/\varepsilon)/\log\log(T/\varepsilon))$.
Applying the Markov bound and standard amplification techniques leads to the query
complexity in the statement of Theorem~\ref{mainthm}.
\end{description}

Our manuscript is organized as follows. In Subsecs.~\ref{step1},~\ref{step2},~\ref{step3},
and~\ref{step4}
 we describe
the simulation of a continuous-time $T$ query algorithm by a
discrete $O(T \log T /\log\log T)$ query algorithm in detail. In
Subsec.~\ref{impcomp} we estimate the amount of full queries as a function of the output fidelity.
Concluding remarks are in Section~\ref{conc}.

\section{Discrete query simulation of continuous-time query algorithms}
\label{sec2}
To obtain a discrete query simulation, we need a discretization of
the continuous-time evolution performed by the algorithm.
For this reason, we define
a \textit{fractional query}  as the operation
\begin{equation}\label{eq:frac-query}
Q_x^{\theta} \ket{j} = e^{-i \theta H_x}\ket{j}= e^{- i \theta x_j}\ket{j},
\end{equation}
for $j \in \{0,1,\ldots,N-1\}$, and its \textit{fractional cost}
is $|\theta|/\pi$. We assume $-\pi < \theta \le \pi$.
When $\theta = \pi$, this is a full query, as defined by
Eq.~(\ref{eq:discrete3}).
A fractional query algorithm alternates driving unitaries and fractional
queries, and its fractional query cost is the sum of the fractional
costs of its queries.

\begin{figure}[!ht]
\vspace{-.4cm}
\begin{equation*}
\Qcircuit @C=.7em @R=.4em{ & \\
&  & \lstick{\ket{+}}    &\ctrl{1} & \gate{R^a_{\theta'}} & \ctrl{1} &\qw &\rstick{\ket{+}}  & \push{\rule{0em}{5em}} \\
& \qw& \qw & \multigate{3}{Q_x} & \qw & \multigate{3}{Q_x} & \qw  \\
& \qw& \qw & \ghost{Q_x} & \qw & \ghost{Q_x} & \qw  \\
& \qw& \qw & \ghost{Q_x} & \qw & \ghost{Q_x} & \qw  \\
& \qw& \qw & \ghost{Q_x} & \qw & \ghost{Q_x} & \qw  \\
}
\end{equation*}
\caption{Simulation of the fractional query $Q_x^{\theta'}$ of Eq.~(\ref{eq:frac-query}) using two full queries $Q_x$ controlled in the state $\ket{1}$ of an ancilliary qubit. The operation $R^a_{\theta'}= \exp(-i \theta' (\one - \sigma_x)/2) $ applies the desired phase to the
state $\ket{-}$ of the ancilla, with $\ket{\pm}=[\ket{0}\pm \ket{1}]/\sqrt{2}$. The operator $\sigma_x$ is the Pauli  bit-flip operator.}
\label{exact}
\end{figure}
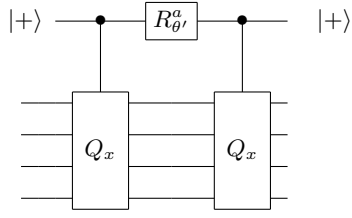

It is straightforward to approximate a continuous-time algorithm
with continuous query time $T$, by a
fractional query algorithm whose total fractional query cost is $T/\pi$ --- but
whose \textit{actual number of fractional queries} $p$ may be much larger
than $T$. Since a fractional query can be easily simulated
using two full queries (Fig.~\ref{exact}), an algorithm that makes $p \gg T$ fractional
queries would be simulated using $2p$ full queries.
This yields an undesirably large overhead for the discrete simulation of the original continuous-time algorithm.
(The problem of simulating fractional powers of arbitrary unitary black-boxes is studied in~\cite{MMS08}).
We introduce another method to approximate fractional queries by full
queries with little loss in efficiency.
What we show is that there is a way of organizing the structure of the
driving and query operations so that many of the full queries
may be omitted with only a small loss in accuracy.  The overall
procedure can be made to succeed with constant probability, and
when it succeeds, the resulting state has very high fidelity to
the state output by the original continuous-time
algorithm.

\subsection{Converting a continuous-time query algorithm to a fractional query algorithm}\label{step1}

This subsection shows how to simulate a continuous-time query algorithm in terms of
a fractional query algorithm that is efficient in terms of its fractional
query cost. We construct this simulation through a straightforward application of
a time-dependent version of the Trotter formula. For arbitrary precision $\varepsilon>0$,
the Trotter-Suzuki approximation allows us to approximate
a continuous-time $T$ query algorithm using fractional queries,
such that the fractional query cost is $T/\pi$.
The construction depends on the \textit{average norm} (or average action)
of the driving
Hamiltonian, defined as
\begin{equation}
r = \frac{1}{T}\int_{0}^{T}\Vert D(t) \Vert\,dt \, .
\end{equation}
Here, $\| \cdot \|$ is the sup-norm defined as $\| H \| = \sup_{\ket{\psi}} \| H \ket{\psi} \|/ \| \ket{\psi} \|$.
We assume that $\| D(t) \|$ is an $L_1$ function,
so that $r$ is well-defined.  (Actually, since we only need $r$ as an
upper bound, it is sufficient that $\| D(t) \|$ is bounded above by an
$L_1$ function.)
Although the number of fractional queries grows proportionally to $r$,
our simulation technique ultimately results in a number of discrete
queries that is independent of the value of $r$.
For fidelity $1 - \varepsilon_1$, it is sufficient to decompose $[0,T]$ into
$p \ge 2 T^2 r/ \sqrt{\varepsilon_1}$ subintervals of size $T/p$.
The fractional query algorithm alternates between evolution under $D(t)$
and evolution under $H_x$ for time $\theta = T/p$.

To handle the case of time-dependent Hamiltonians, we apply an extension,
due to~\cite{HuyghebaertR1990}, of the first-order Trotter product formula
to time-dependent Hamiltonians.
For any time-dependent Hamiltonian $A(t)$, let $U_A(t_b,t_a)$ denote the
unitary operation corresponding to Schr\"{o}dinger evolution under $A(t)$
from $t=t_a$ to $t=t_b$~\cite{unitary-comment}.
Then, by~\cite{HuyghebaertR1990},
\begin{equation}
\pp U_{A+B} (x+\delta,x) - U_A(x+\delta,x) U_B(x+\delta,x) \pp \le
 \!\!
\int_{x}^{x+\delta}
\!\!\!\! \int_{x}^{y} \!\!
\pp [A(y),B(z)]\pp
\,dz
\,dy.
\end{equation}
When $B(t)$ is constant, with $\Vert B \Vert = 1$, this simplifies to
\begin{equation}\label{eq:simptrotter}
\pp U_{A+B} (x+\delta,x) - U_A(x+\delta,x) U_B(x+\delta,x) \pp
\le 2 \delta \int_{x}^{x+\delta}
\Vert A(y) \Vert \,dy.
\end{equation}
In our algorithmic context, we replace $A(t)$ with $D(t) $ and $B(t)$ with $H_x$, and we define the unitaries
\begin{equation}
V_k = U_D((k+1)\theta,k\theta)
\mbox{\hspace*{6mm}and\hspace*{6mm}}
W_k=U_{D+H_x}((k+1)\theta,k\theta).
\end{equation}
Then, by Eq.~(\ref{eq:simptrotter}), for all $k \in \{0,1,\ldots,p-1\}$,
\begin{equation}
\Vert W_k - V_k e^{-i\theta H_x} \Vert
\le 2 \theta \int_{k \theta}^{(k+1)\theta} \Vert D(t)\Vert \,dt \, ,
\end{equation}
from which it follows that
\begin{eqnarray}
\nonumber
\left\Vert
W_{p-1} W_{p-2}\cdots W_0
-
\left(V_{p-1} e^{-i\theta H_x}\right)
\left(V_{p-2} e^{-i\theta H_x}\right)
\cdots
\left(V_0 e^{-i\theta H_x}\right)
\right\Vert
& \le &
2 \theta \sum_{k=0}^{p-1} \int_{k \theta}^{(k+1)\theta} \Vert D(t) \Vert \,dt \\
\nonumber
& = & 2 \theta T r \\
\nonumber
& = & 2 T^2 r/p \\
& \le & \sqrt{ \varepsilon_1}.
\end{eqnarray}
It follows that, if $\ket{\psi_1}$ is the final state of the continuous-time
algorithm, and $\ket{\psi_2}$ is the final state of the approximating fractional
query algorithm, and $p =\lceil 2 T^2 r/ \sqrt{\varepsilon_1} \rceil$ then
\begin{equation}
| \braket{\psi_1}{\psi_2} | \ge \sqrt{1 - \varepsilon_1}.
\end{equation}

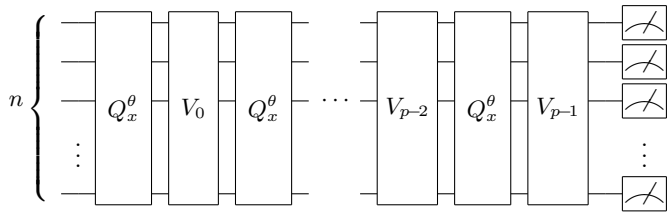
\begin{figure}[!ht]
\begin{equation*}
\hspace{-2.4cm}
\Qcircuit @C=.7em @R=.2em {
&&  \qw  & \multigate{4}{Q_x^\theta}
& \multigate{4}{V_0} &  \multigate{4}{Q_x^\theta}
& \qw & &  & & \multigate{4} {\! V_{p\!-\!2} \!} & \multigate{4} {Q_x^\theta} & \multigate{4}{\! V_{p\!-\!1} \!}
& \qw & \meter\\
&&  \qw   & \ghost{Q_x^\theta}  & \ghost{V_0}
  & \ghost{Q_x^\theta} & \qw & & & &\ghost{\! V_{p\!-\!2} \! }
  & \ghost{Q_x^\theta} & \ghost{\! V_{p\!-\!1} \!}& \qw & \meter\\
\lstick{n \ } && \qw    & \ghost{Q_x^\theta}
&   \ghost{V_0}   & \ghost{Q_x^\theta}
& \qw & & \hspace{-.2cm} \cdots & & \ghost{\! V_{p\!-\!2} \!}   &\ghost{Q_x^\theta}
& \ghost{\! V_{p\!-\!1} \!}& \qw & \meter  \\
 \push{\rule{0em}{2em}}&& \vdots  &   &   &   &      & & & & &   &   &
 & \vdots \\
&& \qw    & \ghost{Q_x^\theta}   & \ghost{V_0}
  & \ghost{Q_x^\theta} & \qw & & & & \ghost{\! V_{p\!-\!2} \!}
&\ghost{Q_x^\theta}& \ghost{\! V_{p\!-\!1} \!}    & \qw &\meter
\gategroup{1}{1}{5}{1}{.5em}{\{}
}
\end{equation*}
\caption{Circuit approximation of a continuous-time $T$ query algorithm.
Each of the $p$ fractional queries $Q_x^\theta$ realize the evolution given by
Eq.~(\ref{eq:frac-query}), with $\theta=T/p \in O(\sqrt{\varepsilon_1}/(rT))$.}
\label{FOC}
\end{figure}

In Fig.~\ref{FOC} we show the $\varepsilon_1$-approximation to the algorithm in the
continuous-time query model. We assume that $V_k$ and $W_k$ act on a set of $n$ qubits (where
$n \ge \log(N)$), but extensions to larger-dimensional systems are
possible.
We refer to these $n$~qubits as the {\em system} to distinguish them from
additional qubits (ancillas) that will be introduced later.
Although the total fractional query cost is $T/\pi$, it should be noted that the total
number of fractional queries
is $T^2 r \sqrt{2/\varepsilon_1}$, which may be much larger than $T$.
(No assumption is
made about the value of $r$.)  For this reason, the full-query
simulation obtained by replacing each fractional query by
the circuit of Fig.~\ref{exact}
may yield an undesirable overhead.  Our full construction will instead result in a
discrete query cost that is $O(T \log T /\log\log T)$, independent of $r$ and $\varepsilon_1$.

\subsection{Simulating fractional queries with low amplitude controlled discrete queries}\label{step2}

\begin{figure}[!ht]
\begin{equation*}
\hspace{-2.4cm}
\Qcircuit @C=1.em @R=.3em {
\lstick{\ket{0}} & \qw & \gate{R_1} & \ctrl{2} & \gate{R_2} & \qw & \meter & \rstick{\left\{ \begin{matrix}  \ket{0}  \otimes e^{i \theta/2} Q_x^\theta \ket{\psi}  ,  \ p_s \ge 1-T/p \ , \\
 \hspace{.15cm} \ket{1} \otimes e^{-i \pi/4} Q_x^{-\pi/2} \ket{\psi} ,  \  p_f \le T/p \ .\end{matrix} \right.} \\
 \\
& \qw & \qw & \multigate{3}{Q_x} & \qw & \qw\\
 \lstick{ \vspace{.1cm}\ket{\psi} \  } & \qw & \qw & \ghost{Q_x} & \qw & \qw \\
& \qw & \qw & \ghost{Q_x}& \qw & \qw \\
& \qw & \qw & \ghost{Q_x} & \qw & \qw \gategroup{3}{3}{6}{1}{.5em}{\{}
}
\end{equation*}
\caption{Probabilistic simulation of the fractional query $Q_x^\theta$ using a single discrete query $Q_x$ controlled on the state $\ket{1}$ of a control qubit. After the measurement,  $Q_x^\theta$ is performed with a success probability $p_s \ge 1-T/p$. Successful simulation occurs if the state of the control qubit is projected into $\ket{0}$.}
\label{FOprob}
\end{figure}
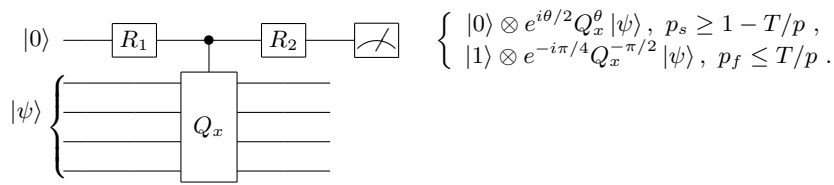

This subsection shows how to replace every fractional query $Q_x^\theta$ in Fig.~\ref{FOC}
by the probabilistic simulation of Fig.~\ref{FOprob}.
Without loss of generality, we may assume $0 \le \theta \le\pi$.
The idea is to add an ancillary (control) qubit initially in $\ket{0}$ and act on it by $R_1$ as
\begin{equation}
\label{r1}
R_1 \ket{0} = \frac{1}{\sqrt{v}} \left[ \sqrt{\cos \theta/2} \ket{0} + i \sqrt{\sin \theta/2} \ket{1} \right] \ ,
\end{equation}
with $v=\cos \theta/2 + \sin \theta/2$.
The full query $Q_x$ is then implemented controlled on the state $\ket{1}$ of the ancilla (i.e., a controlled-$Q_x$ operation), and the ancilla is acted on by $R_2$, given by
\begin{equation}
\label{r2}
R_2 \left\{
\begin{matrix}
\ket{0} & \mapsto& \frac {1}{ \sqrt{v}} [\sqrt{\cos \theta/2} \ket{0} +  \sqrt{\sin \theta/2} \ket{1}] \ ,  \\
\\
\ket{1} & \mapsto & \frac {1} {\sqrt{v}}  [\sqrt{\sin \theta/2} \ket{0} -  \sqrt{\cos \theta/2} \ket{1}] \ .
\end{matrix}
\right.
\end{equation}
Finally, a projective measurement in the computational basis of the ancilla is performed.

To show that the above algorithm implements a probabilistic simulation of $Q_x^\theta$, we write, for all $\theta \in (- \pi, \pi]$,
\begin{equation}
\label{fracoracledef}
Q_x^\theta = e^{-i \theta/2} [\cos(\theta/2)\one + i \sin(\theta/2)Q_x] .
\end{equation}
The ancilla-system state before the measurement is
\begin{equation}
 \frac{1}{v} \left[ e^{i \theta/2} \ket{0} \otimes Q_x^\theta \ket{\psi}+ \sqrt{\sin \theta} e^{-i \pi /4} \ket{1} \otimes Q_x^{-\pi /2} \ket{\psi} \right] \ .
\end{equation}
The measurement in the ancilla projects the state of the system into (up to irrelevant global phase factors) $Q_x^\theta \ket{\psi}$ or $Q_x^{-\pi /2} \ket{\psi}$, with probabilities $p_s= 1/v^2$ and $p_f=1-p_s$, respectively.  Since $\theta=T/p \in O(\epsilon/(rT))$,
we obtain $\sin \theta/2 \le \theta/2=T/(2p)$, and thus $p_s \ge 1-T/p$.

For reasons described in Subsec.~\ref{step4}, our final simulation also requires the probabilistic implementation of the (conjugated) fractional query  $Q_x^{- \theta}$, with similar success probability. We achieve this by replacing $R_1$ with   $R'_1=\sigma_z R_1$ in the circuit of Fig.~\ref{FOprob}. The operator
$\sigma_z$ is the diagonal Pauli operator of the ancilla.
In the event of failure (i.e., if the ancilla state is projected into $\ket{1}$ after measurement), the system state is acted on by the operation $Q_x^{\pi/2}$ up to irrelevant global phase factors. We usually refer to the undesired operations $Q_x^{\pi/2}$ and $Q_x^{-\pi/2}$ implemented in a failed simulation
as {\em errors}.

\subsection{Approximating segments of control qubits by low Hamming weight states}\label{step3}

\begin{figure}[!ht]
\begin{equation*}
\hspace{-0cm}
\Qcircuit @C=.7em @R=.3em {\lstick{\ket{0}}& \gate{R_1}& \qw  & \qw&    \ctrl{9} &\qw& \qw & \qw& \qw & \qw &  & & & \qw& \qw& \qw & \qw & \qw   & \gate{R_2} & \qw &\meter & \\
\lstick{\ket{0}}& \gate{R_1}&   \qw& \qw & \qw &\qw  & \qw & \qw &   \ctrl{8} &\qw  & &  & & \qw& \qw&\qw & \qw & \qw&    \gate{R_2} & \qw & \meter & \rstick{ \ m}\\
\\
\vdots & & & & & & & & & &   & \hspace{-.2cm}\cdots  \\
\\
\\
\lstick{\ket{0}}& \gate{R_1} & \qw& \qw & \qw  & \qw & \qw & \qw & \qw & \qw   & & & & \qw&  \ctrl{3} &  \qw&\qw &\qw  & \gate{R_2} & \qw  & \meter & \\
\\
\\
& & \qw&    \qw & \multigate{3}{\; Q_x \;} & \qw & \multigate{3}{\ V_0 \ } & \qw & \multigate{3}{\; Q_x \;} & \qw & &  & & \multigate{3} {V_{m-2}} & \multigate{3} {\; Q_x \;} & \qw & \multigate{3} {V_{m-1}} & \qw &\qw \\
& & \qw&   \qw & \ghost{\; Q_x \;} & \qw & \ghost{\ V_0 \ } & \qw & \ghost{\; Q_x \;} & \qw & &  && \ghost{V_{m-2}} &\ghost{\; Q_x \;} & \qw & \ghost{V_{m-1}} & \qw& \qw \\
& & \qw&   \qw & \ghost{\; Q_x \;} & \qw & \ghost{\ V_0 \ } & \qw & \ghost{\; Q_x \;} & \qw & &  && \ghost{V_{m-2}} & \ghost{\; Q_x \;} & \qw &\ghost{V_{m-1}}  & \qw& \qw \\
& & \qw&   \qw & \ghost{\; Q_x \;} & \qw & \ghost{\ V_0 \ } & \qw & \ghost{\; Q_x \;} & \qw & & & & \ghost{V_{m-2}}& \ghost{\; Q_x \;} & \qw &\ghost{V_{m-1} }  & \qw& \qw \\
&&&&&&&&&&&&&&&&&& U
\gategroup{1}{22} {7}{22}{.4em}{\}}
\gategroup{1}{4}{13}{17}{1em}{--}
}
\end{equation*}
\caption{The first size-$m$ segment obtained by replacing every $Q_x^\theta$ in Fig.~\ref{FOC} by the probabilistic simulation of Fig.~\ref{FOprob}. If  $m =\lfloor  1/(4\theta) \rfloor$, the total probability of success after measurement  is bounded below by 3/4. Successful simulation occurs if every ancilla is projected into $\ket{0}$ after measurement. The operator $U$ denotes the action induced by the operations inside the dashed box. The overall unitary action before the measurement is  $R_2^{\otimes m} U R_1^{\otimes m}$.}
\label{smallcircuit}
\end{figure}
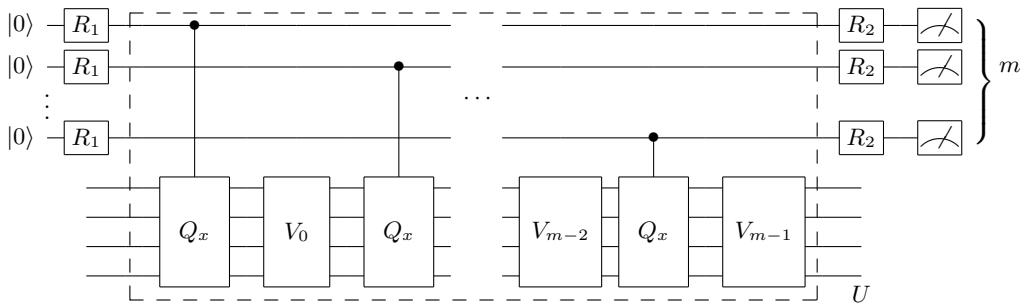

This subsection shows how the low-amplitude controlled discrete queries from the previous subsection
can be efficiently approximated by full queries.
Our construction makes sense for any contiguous segment consisting of $m$ of the controlled discrete queries,
as long as $m \le p$ and $m \in \Omega(1/\theta)$.
For the purposes of the error-correcting procedure in the next subsection, we set $m$ so that
$m \theta = 1/4$~\cite{segmentcomment}. In Fig.~\ref{smallcircuit} we show the first size-$m$ segment ($m$ is the number of full queries) appearing in the original circuit of Fig.~\ref{FOC} after each fractional query has been replaced by its probabilistic simulation, as explained in Subsec.~\ref{step2}.

We begin by observing that, since the state of all $m$ control qubits after the action of the operations $R_{1}$ is
\begin{equation}
\ket{\chi} = R_1^{\otimes m}\ket{0}^{\otimes m} =
\frac{1}{v^{m/2}}\left(\sqrt{\cos\theta/2}\ket{0} + i \sqrt{\sin\theta/2}\ket{1}\right)^{\otimes m} \; ,
\end{equation}
and $m(1/v)\sin\theta/2 \approx 1/8$, the amplitudes of this state are concentrated at the $m$-qubit
states (in the computational basis) with small Hamming weight.
This means that we can approximate $\ket{\chi}$ by a superposition of sufficiently low Hamming weight basis states.
Intuitively, the Hamming weight of the control qubits corresponds to the number of queries actually performed,
suggesting this is a step in the right direction.

To make this approximation precise, let $\Delta(\cdot)$ denote Hamming weight, and define
\begin{equation}
P = \sum_{z \in \{0,1\}^m, \Delta(z) \le k} \ket{z}\bra{z} \; ,
\end{equation}
the projector into basis states $\ket{z}$ of Hamming weight at most $k$.
Also define
\begin{equation}
\ket{\chi'} = \frac{P\ket{\chi}}{\sqrt{\bra{\chi} P \ket{\chi}}}.
\end{equation}
Then, since $|\braket{\chi'}{\chi}|^2$ is the sum of the absolute values squared of all
amplitudes in $\ket{\chi}$ for basis states of Hamming weight up to $k$,
\begin{equation}
|\braket{\chi'}{\chi}|^2 = 1 - \sum_{j > k} \begin{pmatrix} m \cr j \end{pmatrix}(1 - B)^{m-j}B^j,
\end{equation}
where
\begin{equation}
B = \frac{\sin\theta/2}{\cos\theta/2 + \sin\theta/2} = \frac{1}{8m} + O\left(\frac{1}{m^2}\right).
\end{equation}
For asymptotically large $m$ (or $T$), this is essentially the probability that a Poisson distributed
random variable with mean $1/8$ is less than or equal to $k$, which is bounded below by
\begin{equation}
1 - \frac{(1/8)^k}{k!}e^{-1/8}.
\end{equation}
Assuming that the number of segment computations performed is $O(T/\varepsilon_2)$, setting
$\braket{\chi'}{\chi}^2 \ge 1 - \varepsilon_2\varepsilon_3/T$ is sufficient
for the cumulative reduction in fidelity over all the segment computations to
be below $\varepsilon_3$.
To attain this, it is sufficient to set
\begin{equation}
\label{eq:cutoff}
k \in O\left(\frac{\log(T/\varepsilon_2\varepsilon_3)}%
{\log\log(T/\varepsilon_2\varepsilon_3)}\right).
\end{equation}

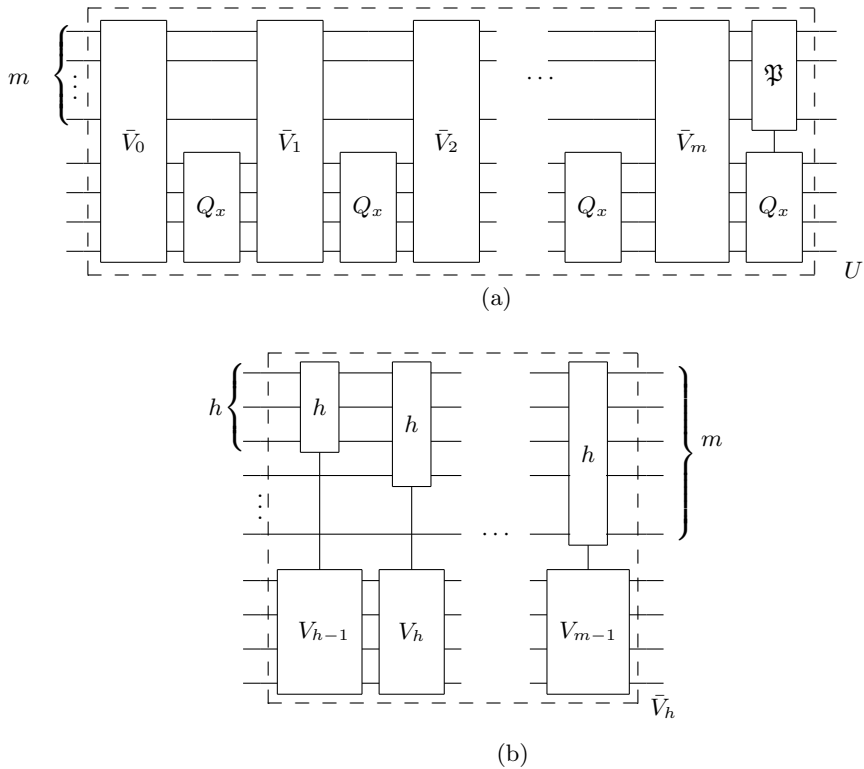
\begin{figure}[!ht]
\begin{equation*}
\hspace{-2.4cm}
\Qcircuit @C=.7em @R=.3em { &\qw& \multigate{12}{ \ \bV_0 \ }  & \qw & \multigate{12}{\ \bV_1 \ } & \qw & \multigate{12}{\ \bV_2 \ }  & \qw & & & & \qw & \qw &\multigate{12}{\ \bV_{ m} \ }  & \multigate{6}{ \mP } \qwx[9] &    \qw     &  \qw \\
 &\qw& \ghost{ \ \bV_0 \ }  & \qw & \ghost{\ \bV_1 \ } & \qw & \ghost{\ \bV_2 \ }  &\qw  & & & & \qw & \qw &\ghost{\ \bV_{ m} \ }  &\ghost{\mP}  & \qw &  \qw  \\
\lstick{m \ \ \ } &   \vdots \ \  & && &  &  & & & & \hspace{-.2cm}\cdots & & & & & & & & \\
 &    \\
 &   \\
 &  \\
 &\qw&  \ghost{ \ \bV_0 \ }  & \qw & \ghost{\ \bV_1 \ } & \qw & \ghost{\ \bV_2 \ }   & \qw & & & & \qw & \qw &\ghost{\ \bV_{ m} \ }  & \ghost{\mP} & \qw &  \qw \\
& & \\
& & \\
& \qw& \ghost{ \ \bV_0 \ } & \multigate{3}{Q_x} & \ghost{\ \bV_1 \ } & \multigate{3}{Q_x} & \ghost{\ \bV_2 \ } & \qw &  & & & \multigate{3} {Q_x} & \qw &\ghost{\ \bV_{ m} \ }  & \multigate{3}{Q_x} &\qw& \qw \\
&\qw&  \ghost{ \ \bV_0 \ } & \ghost{Q_x} & \ghost{\ \bV_1 \ } & \ghost{Q_x} & \ghost{\ \bV_2 \ } & \qw & & & &\ghost{Q_x} & \qw & \ghost{\ \bV_{ m} \ }   & \ghost{Q_x} & \qw& \qw \\
&\qw&   \ghost{ \ \bV_0 \ }& \ghost{Q_x} & \ghost{\ \bV_1 \ } & \ghost{Q_x} & \ghost{\ \bV_2 \ } & \qw &  & & & \ghost{Q_x} & \qw & \ghost{\ \bV_{ m} \ }  & \ghost{Q_x} & \qw& \qw \\
&\qw&   \ghost{ \ \bV_0 \ } & \ghost{Q_x} & \ghost{\ \bV_1 \ } & \ghost{Q_x} & \ghost{\ \bV_2 \ }  & \qw & & & & \ghost{Q_x} & \qw & \ghost{\ \bV_{ m} \ }  & \ghost{Q_x} & \qw&\qw \\
&&&&&&&&&&&&&&&&&U
\gategroup{1}{3}{13}{15}{1em}{--}
\gategroup{1}{1}{7}{1}{.4em}{\{}
\\
\\
\\
\\
&&&&&&& \mbox{(a)} \\
\\
\\
\\
\\
}
\end{equation*}
\begin{equation*}
\hspace{-2.4cm}
\Qcircuit @C=.7em @R=.5em{   & \qw & \multigate{2}{ h } \qwx[8] &  \multigate{3}{ h } \qwx[8]  & \qw & & & &   &  \multigate{6}{ h } \qwx[8]   & \qw & \qw & \\
\lstick{h \ } & \qw & \ghost{ h } &  \ghost{ h }& \qw & & & &  &  \ghost{ h } & \qw & \qw &\\
 & \qw  &\ghost{ h } & \ghost{ h } & \qw & & & &  &  \ghost{ h } & \qw & \qw & \rstick{ \ m} \\
 & \qw  &\qw & \ghost{ h } & \qw & & & & & \ghost{ j } & \qw  & \qw& \\
& \vdots &\\
& \\
 & \qw & \qw & \qw & \qw & &  \cdots& & & \ghost{ j } & \qw & \qw &
& \\
& \\
& \qw  & \multigate{3}{\ V_{h -1 }} & \multigate{3}{\; V_{h} \; } & \qw & & & &  & \multigate{3}{V_{m-1}  } & \qw& \qw  \\
& \qw  &\ghost{\ V_{h-1  }} &\ghost{\; V_{h} \; } & \qw & & & &  & \ghost{ V_{m-1} } & \qw & \qw \\
& \qw   & \ghost{\ V_{h-1  }} &\ghost{\; V_{h} \; }  & \qw &  & & & & \ghost{V_{m-1}  } & \qw & \qw\\
& \qw  & \ghost{\ V_{h -1 }} &\ghost{\; V_{h } \; } & \qw &  & & & & \ghost{V_{m-1}  } & \qw & \qw\\
& & & & & & &  & & & & \bar{V}_h
\gategroup {1}{1}{3}{1}{.7em}{\{}
\gategroup{1}{3}{12}{10}{.7em}{--}
\gategroup{1}{13} {7}{13}{.4em}{\}}
\\
\\
\\
\\
&&&&&&& \mbox{(b)} \\
\\
\\
}
\end{equation*}
\caption{(a) Equivalent quantum circuit for the implementation of the circuit $U$ within the dashed box in Fig.~\ref{smallcircuit}. The last query  is controlled on the parity $\mP$ of the state of the $m$  control qubits, depending on whether $m$ is odd or even.
(b) Description of the unitary $\bV_h$.
Each operation $V_{h-1},\cdots,V_{m-1}$ is controlled on the Hamming weight of the ancilliary state, enclosed by the corresponding boxes,  being $h$. We extend the notation so that $V_{-1}=\one$.}
\label{eqsc}
\end{figure}

Although changing $\ket{\chi}$ to $\ket{\chi'}$ \textit{suggests that} the number of controlled full
queries can be reduced to $k$, the circuit in Fig.~\ref{smallcircuit} must be rearranged
to make this possible (as it is, it makes $m$ queries regardless of $k$).
The idea is to replace the controlled-queries interleaved with driving unitary operations
in Fig.~\ref{smallcircuit} with an equivalent circuit composed with \textit{fixed} discrete queries
interleaved with \textit{controlled} driving unitaries.
The form of the revised circuit is illustrated in Fig.~\ref{eqsc}.

The action of the gates $\bV_0, \ldots, \bV_{m}$ is defined as follows.
Let $i_{h}$ be the position of the $h$-th $1$ in a state of the $m$ control qubits, in the computational basis. (These states form a complete orthogonal basis allowing a case-by-case analysis for the equivalence.) The positions range from $0$ (top) to $m-1$ (bottom).
Then $\bV_0$ is controlled by the state of the first $m$ qubits and
acts as follows: if $i_{1}=0$ it does nothing. Otherwise
 it applies the sequence $V_{0}, \ldots, V_{i_1-1}$, unless $i_1$ is not well-defined
(that is, if the state of the $m$ control qubits is $\ket{0}^{\otimes m}$), in which case it applies all the unitaries $V_0, V_1, \ldots, V_{m-1}$ in the segment.
For $h > 0$, $\bV_h$ applies $V_{i_{h}}, \ldots, V_{i_{h+1}-1}$ if $i_{h}$ and $i_{h+1}$ are well-defined;
does nothing if $i_{h}$ is not well-defined; and applies $V_{i_{h}}, \ldots, V_{m-1}$ if
just $i_{h+1}$ is not well-defined.
It is easy to see that this circuit exactly simulates the one in Fig.~\ref{smallcircuit}. It still makes $m \in O(rT/\sqrt{\varepsilon_1})$ full queries.

Note that if the control qubits are in a state of Hamming weight smaller than $h$ then $\bV_h, \ldots, \bV_{m}$
do nothing and can be removed from the circuit.
This follows from the above description;
it can also be deduced by noting that the action of $\bV_{h}$ depends on the Hamming weight of the state
of the control qubits in the manner shown in Fig.~\ref{eqsc}(b).  Full queries square to the identity: $Q_x^2 = \one$.
For this reason,  it is possible to truncate
the last $m-k$ queries of the circuit specified in Fig.~\ref{eqsc}(a) without changing its effect on superpositions of basis states of Hamming weight bounded by $k$; if $m-k$ is odd, we need to change the control of the parity-controlled query at the end. For $k$ chosen as in Eq.~(\ref{eq:cutoff}), the truncated circuit involves $k+1 \in O(\log(T/\varepsilon_2\varepsilon_3)/\log\log(T/\varepsilon_2\varepsilon_3))$
full queries (which is much less than $m$).
It outputs a $O(\varepsilon_2\varepsilon_3/T)-$approximation to the state output by $U$ in Fig.~\ref{smallcircuit}.  We apply this truncation to all size-$m$ segments in the circuit. The low query cost as a function of $T$ from the truncation motivates the error correcting procedure that we now explain.


\subsection{Correcting erroneous fractional queries}\label{step4}

This subsection explains how to correct the erroneous $Q_x^{\pm \pi/2}$ queries that occur
when the measurements in Subsec.~\ref{step2} fail.

As mentioned in Subsec.~\ref{step3}, the computation is divided into $m$ segments so that $m \theta = 1/4$. Note that there are $4T$ such segments. Before the approximation of the previous Subsection is made, we have:
(a) the probability that a size-$m$ segment is successfully completed is at least $3/4$;
(b) conditional on a segment completing unsuccessfully, the expected number of errors is
upper bounded by $1/4$.
Property (b) was obtained by considering that the error probability per control qubit is bounded by $\theta=T/p$ (Subsec.~\ref{step2}), and there are $m = 1/(4\theta)$ control qubits per segment.

We now describe an error correction procedure for each segment that succeeds with an expected number of extra segments that is bounded by a constant. Each of these extra segments will be ultimately simulated using the truncation explained in Subsec.~\ref{step3}.
With this approximation, the expected number of queries is proportional to the cost of the original segment computation, namely $O(\log(T/\varepsilon_2\varepsilon_3)/\log\log(T/\varepsilon_2\varepsilon_3))$.

The following analysis is valid for the {\em exact} case, without invoking the approximation of Subsec.~\ref{step3}.
Intuitively, the errors between the two approximations accumulate linearly,
which is shown rigorously in the next section.

\begin{figure}[hbt]
\includegraphics*[width=4.in]{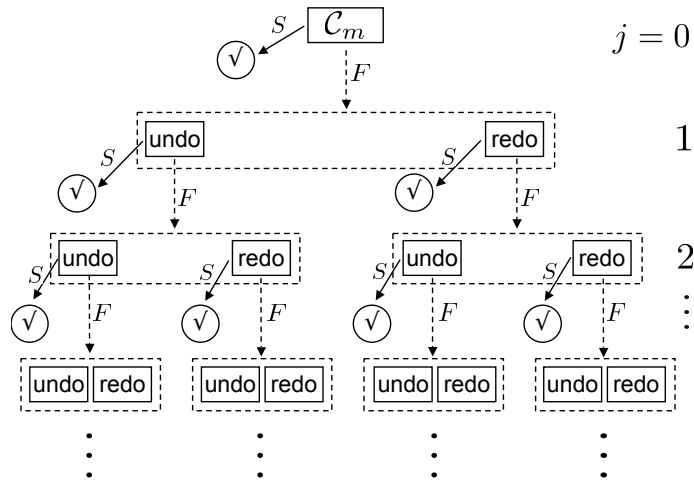}
\caption{Branching process for the iterative error correcting step. We let ${\cal C}_m$ denote a size-$m$ segment  of Fig.~\ref{FOC} to be simulated probabilistically as described in Subsec.~\ref{step2}. When any simulation fails (F) we attempt to correct it by undoing the failed circuit and redoing it with the right operations.  The undoing and redoing circuits are size-$m'$ circuits, with $m' < m$ (they require $m'$ operations $Q_x^{\pm \theta}$). These are also implemented probabilistically with $m'$ controlled full queries (Subsec.~\ref{step2}), yielding a success probability bounded below by $3/4$.  The dashed boxes and arrows are associated with the nodes and branches of the tree, respectively. Successful simulation (S) occurs when ${\cal C}_m$ is simulated successfully; that is, when the undoing and redoing circuits associated to all the visited nodes are simulated successfully (check marks). Variable $j$ denotes the level of the branching process. }
\label{branch}
\end{figure}

First, note that, whenever erroneous fractional queries occur, it is known from the ancilla
measurements in exactly what positions the resulting errors are.
Since errors are unitary operations, it is then possible to \textit{undo} the entire segment, and then \textit{redo} it.
At a high level, the undoing operation is implemented by simulating the fractional queries
and the interleaved driving unitaries in reverse while inserting the $Q_x^{\pm \pi/2}$ in place
of $Q_x^{\mp \theta}$ wherever an error has occurred (this aspect is described in more detail further
below).
The undo and the redo each succeed with probability at least $3/4$, but they may each fail.
If the undo or redo step fails, we iterate the recovery procedure.
For instance, if the undo step fails, we must undo the failed undo step, and then redo the failed undo step.
If these two actions succeed, we can continue with the original redo step from the recovery procedure.
Success occurs when all the recovery steps are successfully implemented.
Figure~\ref{branch} illustrates the error correction process for an original segment.

For the implementation of a segment, there are several types of segment-like computations related to it: the \textit{original segment}, segments corresponding to undo operations for sets of error positions, and the
recursive versions of these (such as the undo operations related to each unsuccessful undo).
We refer to each of these as \textit{segments}.

Our first observation is that the expected number of segments that are computed in order to correctly
compute an original segment is at most $2$.
To see why this is so, note that the branching process for the error correction in Fig.~\ref{branch} can be viewed as a classical random walk on a line, that moves one
step to the right whenever a segment computation succeeds ($S$) and one step to the left whenever a segment
computation fails ($F$). (Upon failure, two segment computations are required: the failed segment computation
must be undone and then redone.)
The random walk starts in the state corresponding to the original segment and the goal is to advance
one step to the right of this state.
Since each of the segment computations succeeds with probability at least $3/4$, this is a biased random
walk with average speed (to the right) bounded below by $3/4-1/4=1/2$. The expected number of steps, or segment computations, is then $A \le 2$.

Although the expected number of segment computations for each original segment ${\cal C}_{m}$ is bounded by a constant, we must take into account that not all segment computations have the same cost. The undo segments become more expensive as the number of errors being corrected increases.
That is, for each erroneous fractional query, the undo segment has to correct all the $Q_x^{\pm \pi/2}$
operations obtained in the failed simulation.  In order to approximate the segment using the truncation procedure explained in Subsec.~\ref{step3},
we absorb each error $Q_x^{\pm \pi/2}$ into the adjacent $V_i$ unitaries.  Once we perform the transformation and truncation from Subsec.~\ref{step3}, a single error then gets multiplied into up to $k$ operations $Q_x^{\pm \pi/2}$, with $k$ as in Eq.~(\ref{eq:cutoff}).
Each of these operations is implemented using two full queries as shown in Fig.~\ref{exact}, replacing $\theta'$ by $\pm \pi/2$. Therefore, with the approximation in mind, for each error obtained in a failed simulation it requires $O(\log(T/\varepsilon_2\varepsilon_3)/\log\log(T/\varepsilon_2\varepsilon_3))$
\textit{additional} full queries in the undo segment, as well as in each of the recursive undo and
redo operations that occur if this segment fails.

We return to the exact case.  Let $C_0$ denote the expected number of operations $Q_x^{\pm \pi/2}$ needed to fix all the errors that occur in all segment computations of the branching process,
starting from the original segment ${\cal C}_{m}$.
For each integer $\alpha \ge 0$, let $q_\alpha$ denote the probability that the initial computation of the original
segment results in $\alpha$ errors ($q_0 \ge 3/4$). As $m$ gets large, $q_{\alpha}$ approximates a Poisson distribution with mean bounded by $m \theta =1/4$.
Also, for each $\alpha \ge 1$, let $C_\alpha$ denote the expected number of operations $Q_x^{\pm \pi/2}$ required
to successfully undo the $\alpha$ errors after a failed computation of the original segment.
Since these $\alpha$ errors will be part of every segment that is associated with this undo operation
and the expected number of such segments is $A$, $C_\alpha \le C_0 + \alpha A$; the $C_0$ term denotes
the expected number of new errors introduced during the undo operation.  If we happen to have $\alpha > 0$ errors in a segment, the expected number of operations $Q_x^{\pm \pi/2}$ we need to perform to correct it is then $C_\alpha$ for the undo step, plus $C_0$ for the redo step.
Considering all the possible outcomes of a segment simulation, we have
\begin{eqnarray}
C_0 & = & q_0\cdot 0 + \sum_{\alpha=1}^{m} q_\alpha(C_\alpha + C_0) \\
& \le & \sum_{\alpha=1}^{m} q_\alpha((C_0 + \alpha A) + C_0) \\
& \le & 2C_0\left(\sum_{\alpha=1}^{m} q_\alpha \right) + A\left(\sum_{\alpha=1}^{m} \alpha q_\alpha\right) \\
& \le & (1/2)C_0 + 2(1/4)
\end{eqnarray}
(where we have used the fact that the expected number of errors is upper bounded by $1/4$),
which implies that $C_0 \le 1$.

In summary, using the error correction procedure, the expected number of segment computations and the expected number of operations $Q_x^{\pm \pi/2}$ needed in the exact case are both $O(T)$.  Using Markov's inequality, the probability of successfully terminating the error-correcting procedure can
be lower bounded by $1 - \varepsilon_2$ by running the procedure for $O(T/\varepsilon_2)$ segment computations and operations $Q_x^{\pm \pi/2}$.  That is, we set a cutoff for the total number of queries used, and terminate the algorithm if this quota is exceeded.  The probability of this happening is at most $\varepsilon_2$.  Note that if we have a classical input to the algorithm, we can just set $\varepsilon_2$ equal to a constant.  If the error correction procedure fails, we start over from the beginning; we can then attempt to run the algorithm $O(\log(1/\varepsilon'_2))$ times, and the probability that one or more attempts succeed is at least $1-\varepsilon'_2$.

If we use the truncation of Subsec.~\ref{step3} to approximate each segment computation, the expected number of full queries is
$O(T \log(T/\varepsilon_2\varepsilon_3)/\log\log(T/\varepsilon_2\varepsilon_3))$, and the probability that the error correction procedure terminates is at least $1-\varepsilon_2$ if we make $O(T \log(T/\varepsilon_2\varepsilon_3)/\varepsilon_2 \log\log(T/\varepsilon_2\varepsilon_3))$ queries.
In the next Subsection, we give a rigorous proof that the resulting final state of the computation has fidelity at least
$\sqrt{1 - 3(\varepsilon_1 + \varepsilon_2 + \varepsilon_3)}$ with the final
state of the continuous-time algorithm, which completes the proof of Thm.~\ref{mainthm}.

\subsection{Rigorous analysis of fidelity}
\label{impcomp}

From subsection~\ref{step1}, we have
$\braket{\psi_1}{\psi_2} \ge \sqrt{1 - \varepsilon_1}$,
where $\ket{\psi_1}$ is the output state of the continuous-time
algorithm and $\ket{\psi_2}$ is the output state of the fractional
query algorithm.  (Assume throughout this section that the overall phases of
the various states have been adjusted to make all inner products positive.)

Consider the algorithm with error correction as described in subsection~\ref{step4},
assuming that the control qubits are in the exact state
$\ket{\chi}^{\otimes O(T/\varepsilon_2)}$ (that is, there is no approximation of
this state by $\ket{\chi'}^{\otimes O(T/\varepsilon_2)}$).
This algorithm can be expressed via purification as a unitary operation whose input state
includes the control qubit state $\ket{\chi}^{\otimes O(T/\varepsilon_2)}$
and whose output state is of the form
\begin{equation}
\sket{\tilde{\psi}_3} = \sqrt{1 - \varepsilon_2}\ket{\psi_2}\ket{0}\ket{g_0}
+ \sqrt{\varepsilon_2}\ket{\psi'}\ket{1}\ket{g_1},
\end{equation}
where the second register tells us if the error correction procedure succeeded.
Note that, if we define $\sket{\tilde{\psi}_2} = \ket{\psi_2}\ket{0}\ket{g_0}$
then we have
\begin{equation}
\braket{\tilde{\psi}_2}{\tilde{\psi}_3} \ge \sqrt{1 - \varepsilon_2}.
\end{equation}
Finally, if we change the control qubit input state to the unitary operation
from $\ket{\chi}^{\otimes O(T/\varepsilon_2)}$ to
$\sket{\chi'}^{\otimes O(T/\varepsilon_2)}$
and define the resulting output state as
$\sket{\tilde{\psi}_4}$ then
\begin{eqnarray}
\braket{\tilde{\psi}_3}{\tilde{\psi}_4}
& = & (\braket{\chi}{\chi'})^{\otimes O(T/\varepsilon_2)} \\
& \ge & \sqrt{1 - \varepsilon_3}.
\end{eqnarray}
Combining these results yields
\begin{equation}
\braket{\tilde{\psi}_1}{\tilde{\psi}_4} \ge \sqrt{1 - 3(\varepsilon_1 + \varepsilon_2 + \varepsilon_3)},
\end{equation}
where $\sket{\tilde{\psi}_1} = \ket{\psi_1}\ket{0}\ket{g_0}$.
This implies that the fidelity between $\ket{\psi_1}$ and the
corresponding portion of $\sket{\tilde{\psi}_4}$ is $1 - \varepsilon$
if $\varepsilon_1 = \varepsilon_2 = \varepsilon_3 = \varepsilon/9$.

\section{Concluding Remarks}
\label{conc}
We have described an efficient discrete-query simulation of a continuous-time query algorithm. For total evolution time $T$ and arbitrary precision, our algorithm requires $O(T \log T / \log\log T)$ full queries and $O(T^2 \log T)$ known unitaries for its implementation, and its probability of success is bounded below by an arbitrary constant.
We expect that the known operation complexity can be reduced; however we do not do so here.

As a consequence, lower bounds on the (discrete) quantum query complexity for a function also are lower bounds on the continuous query complexity, possibly lower by a sub-logarithmic factor.  We note that one can use this simulation to show that the bounded-error query complexity lower bound one would obtain by the polynomial method applied to the quantum query model is also a lower bound for the fractional or continuous query model, without the loss of a $\log T / \log\log T$ factor. It is conceivable, of course, that a continuous query algorithm might be able to achieve this lower bound when a discrete query algorithm cannot.

One way to to show that the polynomial method does not lose a factor of $\log T / \log\log T$ is as follows.  If instead of breaking the algorithm up into blocks of size $m$, we run the algorithm to completion without performing any error correction, the total number of full queries used (after truncation) is now just $O(T)$ rather than $O(T \log T / \log\log T)$.
We can thus use the polynomial method to show that the final amplitudes (after
truncation) are polynomials of degree in $O(T)$ in the variables $x_0$, $x_1$, $\ldots$\, $x_{N-1}$.
If, conditional on the final state of each control bit, we perform the mathematical operation that effectively {\em fixes} things, then this would yield the correct answer for all (low Hamming weight) values of the control bits, and doesn't increase the degree of the amplitude polynomials. Even though these aren't physical operations, they preserve the norm, and thus we get valid approximating polynomials with the same degree upper bounds.
One way to {\em fix} things is to use variables $X_{i,j}$, where $X_{i,j}$ is the result of querying bit $i$ on query number $j$.  The mathematical correction can be to map $X_{i,j} = x_i$ if control bit $j$ is $0$, and to map $X_{i,j} = 1-x_i$ if control bit $j$ is $1$.

There remain a number of open questions about possible improvements.
The total number of queries could possibly be reduced further.  The minimum possible number of full queries is $\Omega(T)$, since a quantum query algorithm with $T$ full queries can be simulated by continuous-time algorithm running for time $\Theta(T)$.  It might be possible to eliminate the factor $\log T/\log\log T$ from the query complexity of our simulation to obtain a tight equivalence between discrete and continuous query algorithms.  In our simulation, this factor arises from the need to break up the algorithm into small segments, each of which can only have small error.  However, without breaking up the algorithm in this way, too many errors accumulate due to failures of the probabilistic simulations for us to correct.  Therefore, it appears that a new way of handling the error correction is needed if we wish to remove the extra factor of $\log T/\log\log T$.

Also, for a computation on an arbitrary initial quantum state (where fast amplification by trivially
repeating the computation cannot be carried out) a failure probability bound of $\varepsilon$ requires
a factor of $\varepsilon^{-1}$ in the query complexity (by our approach, using the Markov bound).
Since the branching process in Section~II D becomes extinct exponentially fast in the generation,
we conjecture that this scaling in $\varepsilon$ can be improved towards $O(\log(1/\varepsilon))$.
Such an improvement may be useful in some settings.

Finally, it is an interesting question to see if the number of driving unitary operations can be reduced
to correspond to the cost of implementing the evolution of the driving Hamiltonian alone (in some
reasonable sense).
In the most general case, with a rapidly varying and strong driving Hamiltonian $D(t)$, we do not expect a considerable reduction: a general $D(t)$ corresponds to a complicated unitary circuit.
However, for better behaved $D(t)$, a reduction is expected. The case where $D(t)$ is time-independent is particularly interesting, and could have relationships with improved Trotter-Suzuki formulas.
In this case, all operations $V_j$ in Fig.~\ref{FOC} are identical and the $R_1$ gates from Fig.~\ref{smallcircuit} and $\bV_h$ gates from Fig.~\ref{eqsc} can all be done using only $O(T\, \mbox{polylog}\ T)$ unitaries.
Unfortunately, we do not know how to also reduce the number of $R_2$ gates, so it remains an open problem
whether the number of unitaries can be reduced to $O(T\, \mbox{polylog}\ T)$ even in the case of a constant
driving Hamiltonian.


\acknowledgments
We thank P. H\o yer and E. Knill for discussions.
This research was supported by Perimeter Institute for Theoretical
Physics, by the Government of Canada through Industry Canada, by
the Province of Ontario through the Ministry of Research and
Innovation, NSERC, DTO-ARO, CFI, CIFAR, CRC, OCE, QuantumWorks, and MITACS. The figures have been made using Q-circuit, available online at
http://info.phys.unm.edu/Qcircuit/.



\begin{thebibliography}{10}
\bibitem{unitary-comment}
$U_A(t_b,t_a)$ can be defined in terms of the recurrence
$U_A(t_b,t_a)=\one - i \! \int_{t=t_a}^{t_b} \! A(t)U_A(t,t_a)\,dt$,
where the operator $\one$ is the identity (no-action) operation.
See~\cite{Messiah1999} for details.
\bibitem{segmentcomment}
If the $m$ obtained is not integer, we can set $m=\lfloor \frac1 {4\theta} \rfloor$.
\end{thebibliography}
\end{document}